\documentclass[fleqn,twoside,twocolumn,nofootinbib,showkeys]{revtex4} 
\usepackage[nocpr]{ujp} 

\begin{document}
\title[Thermal Conductivity of Molecular Crystals]
{THERMAL CONDUCTIVITY OF MOLECULAR\\ CRYSTALS WITH SELF-ORGANIZING DISORDER}
\author{A.I.~Krivchikov}
\affiliation{B.~Verkin Institute for Low Temperature Physics and
Engineering, Nat. Acad. of Sci. of Ukraine}
\address{47, Lenin Ave., Kharkiv 61103, Ukraine}
\email{krivchikov@ilt.kharkov.ua}
\author{G.A.~Vdovychenko}%
\affiliation{B.~Verkin Institute for Low Temperature Physics and
Engineering, Nat. Acad. of Sci. of Ukraine}%
\address{47, Lenin Ave., Kharkiv 61103, Ukraine}%
\author{O.A.~Korolyuk}
\affiliation{B.~Verkin Institute for Low Temperature Physics and
Engineering, Nat. Acad. of Sci. of Ukraine}
\address{47, Lenin Ave., Kharkiv 61103, Ukraine}
\author{O.O.~Romantsova}%
\affiliation{B.~Verkin Institute for Low Temperature Physics and
Engineering, Nat. Acad. of Sci. of Ukraine}%
\address{47, Lenin Ave., Kharkiv 61103, Ukraine}%
\udk{539}  \razd{\secvii}

\autorcol{A.I.\hspace*{0.7mm}Krivchikov,
G.A.\hspace*{0.7mm}Vdovychenko, O.A.\hspace*{0.7mm}Korolyuk et al.}

\setcounter{page}{319}%

\begin{abstract}
The thermal conductivity of some orientational glasses of protonated
$C_2H_5OH$ and deuterated $C_2D_5OD$ ethanol, cyclic substances
(cyclohexanol $C_6H_{11}OH$, cyanocyclohexane $C_6H_{11}CN$,
cyclohexene $C_6H_{10}$), and freon 112 $(CFCl_2)_2$ have been
analyzed in the temperature interval 2--130~K. The investigated
substances demonstrate new effects concerned with the physics of
disordered systems. Universal temperature dependences of the thermal
conductivity of molecular orientational glasses have been revealed.
At low temperatures, the thermal conductivity exhibits a universal
behavior that can be described by the soft potential model. At
relatively high temperatures, the thermal conductivity has a smeared
maximum and than decreases with increase in the temperature, which occurs
typically in crystalline structures.
\end{abstract}
 \maketitle

\section{Introduction}

Orientational glasses form a relatively small class of crystalline
molecular materials, in which the centers of molecular masses reside
at the lattice sites and their orientations and/or conformations are
disordered.\,\,According to the position of the centers of molecular
masses, these substances are crystals; on the other hand, they are
glasses possessing specific properties.\,\,Of special interest are
the temperature dependences of the thermal conductivity $\kappa(T)$
of these objects.\,\,They deviate from the corresponding dependences
typical of orientationally-ordered molecular crystals: there is no
phonon maximum in most of them [1--3].\,\,The dependences
$\kappa(T)$ also differ from those typically observed in structural
glasses; they have no plateau and do not increase with a rising
tem\-perature [4].

It was demonstrated with protonated ethanol [1] that the thermal
conductivity curves of orientational and structural glasses having
basically different molecular disorderings were unexpectedly close
in value and temperature dependence. This finding was then
confirmed in the investigation of the thermal conductivity of
deuterated ethanol [2]. The behavior of the thermal conductivity of
ethanol glasses suggests that the acoustic phonon scattering in
glasses is caused mainly by the orientational molecular disorder
rather than the structural one. This stimulated a search for
common regularities in the dependence $\kappa(T)$ of orientational
glasses.

\section{Materials}

Solid alcohols are suitable objects to investigate the low-temperature anomalies and the thermal properties of disordered
systems [5]. By selecting the thermal prehistory of samples, these
molecular H-bonded solids can be transformed easily into a different
solid state with a structural or orientational disorder. In the
homologous series of monoatomic alcohols, only ethanol (protonated or
deuterated) can exist in the state of orientational glass. This
state is obtainable in both protonated and deuterated ethanols at the
cooling below $T_{g}=97$~K. The obtained glass has a bcc
structure. The investigation of the thermal conductivity of the
orientational glasses of monoatomic alcohols was logically continued
on cyclic alcohols~-- cyclohexanol C$_6$H$_{11}$OH and its analogs:
cyanocyclohexanol C$_6$H$_{11}$CN, cyclohexene C$_6$H$_{10},$ and
freon 112 (CFCl$_2)_2$.

In this class of substances, the state of orientational glass
evolves due to the orientational disorder which occurs, when the
rotational motion of the molecules in a plastic crystal is frozen,
and the molecule can vary its conformation in a flexible carbon
skeleton.

The cyclohexanol molecule is a ring formed by a cyclic carbon
radical with a functional hydroxyl group -OH fixed on it. Structural
and calorimetric investigations show that the pseudospheric form and
the functional group enable the molecule to change into different
conformations and to exhibit an interesting polymorphic behavior
[6--10]. At low temperatures, the substance can exist in several
solid phases depending on the thermodynamic prehistory of the
sample.

Phase I is a stable orientationally disordered state, in which the
molecules rotate almost freely at the sites of the fcc lattice.
Phase I develops due to the crystallization of a liquid at the cooling
below the melting temperature $T_{m}=299$~K. Note that phase I is
supercooled readly at moderate cooling rates and transforms into an
orientational glass at $T_{g}=148$--150~K [9, 11].

Phase II having a tetragonal structure (space group $P\bar{4}2_{1}c$
[8]) can be obtained either at the smooth cooling of phase I below
$T=265.5$~K or at the heating of metastable ordered phase III to
$T=220$--240~K.

Phase III (space group $Pc$) can be obtained by heating an
orientational glass above the glass formation temperature at
$T\approx200$~K. According to recent results [8], the cyclohexanol
ring takes a ``chair'' conformation in all ordered phases.

According to recent findings, there are two factors responsible
for the disorder in the orientational glass of molecular cyclic
structures such as cyclohexanol and the related substances. These
are a mixture of conformations and a disordered orientation of the
molecule as a whole [6].

Cyanocyclohexane C$_6$H$_{11}$CN is also rich in polymorphism due to
the presence of various conformers. In addition to the ring
conformations determined by the position of the cyanic group with
respect to the ring, the cyanocyclohexane molecule has axial and
equatorial conformations. According to IR and Raman data [12], the
latter conformations can interconvert at energies $\sim $$E/k_{\rm
B}=4500$~K. They have identical occupancies in phase I and in the
liquid state; low-temperature orientationally ordered phase II has
only an axial conformation [12].

Orientationally disordered fcc phase I of cyanocyclohexane evolves
due to the crystallization of the liquid ($T_{m}=285$~K). The glassy state
is formed at the cooling of phase I at $T_g\sim 135$ K (see calorimetric
and dielectric spectroscopy data [13, 14]). A small jump of the heat
capacity was also observed near 55~K [15], which was attributed to
the frozen interconversion of the axial and equatorial conformers.
At the heating above 271~K, phase I transforms into stable
orientationally ordered phase II. The conclusion about the state of
phase II of cyanocyclohexane is based only on the measurement of
enthalpy variations during the II-I phase transition [15]. No
structural evidence is available to support it.

Of all the discussed substances, cyclohexene possesses the widest
range of polymorphism. It can be obtained in five solid states
depending on the temperature prehistory of the sample. Owing to the
double bond in the carbon ring, the ``half-chair'' conformation is
the most stable form of the molecule.

Cooling below the crystallization temperature transforms a liquid
sample into a plastic crystal (phase I) having the fcc structure and
a dynamic orientational and conformational disorder. At the fast cooling
below $T_g=81$~K, the plastic crystal changes into an orientational
glass.

Annealing at $T=120$--140~K transforms phase I into stable
completely ordered low temperature phase II. The molecules of phase
II are orientationally ordered.

At the slow cooling below 112 K, phase I transforms into metastable
phase III, in which half the molecules are orientationally
ordered, and the others have dynamic orientational and
conformational disorder. At the further cooling below 83~K, the motion
of the disordered molecules is frozen, and a state, in
which only half the molecules are orientationally ordered, is formed.
Metastable phase III is monoclinic with eight molecules per unit
cell. It exhibits a pseudocubic behavior inherited from the
structure of phase I.

Freon 112 (${\rm CFCl}_2)_2$ is the final substance in the
investigated series of orientational glasses. Its molecule is not
cyclic and has no hydrogen bonds. The orientational glass state
evolves, because the molecules can exist in two molecular
conformations -- {\it gauche} and {\it trans}. The change into an
orientational glass occurs at the cooling of a high-temperature
orientationally disordered bcc phase slightly below $T_g=90$ K.
There is information in the literature [16, 17] that freon 112 is
the most fragile substance among the known orientational glasses.

\section{Results and Discussion}

The thermal conductivities of some orientation glasses (protonated
and deuterated ethyl alcohols, cyclohexanol, cyanocyclohexane,
cyclohexene, and freon 112) have been analyzed. The goal was to
search for regularities in the behavior of the temperature
dependence of the thermal conductivity at low temperatures. The
thermal conductivity of these substances was measured previously
[1--4, 18] in the interval from 2~K to the glass transition
temperature, by using the steady-state method [19, 20].

The temperature dependences of the thermal conductivity of the
orientational glasses of protonated C$_2$H$_5$OH and deuterated
C$_2$H$_5$OD ethanol [1, 2] are illustrated in Fig.~1.

The curves $\kappa(T)$ have three distinct temperature intervals, in
which the thermal conductivity exhibits different behavior: a low
temperature region below 5~K, an interval of 5--50~K, and high
temperature region above 50~K. At low temperatures ($T<5$~K), the
thermal conductivity grows most rapidly. In the interval 5--50~K, it
forms a smeared ``plateau,'' and its growth is weaker than in the
region $T<5$~K. In these intervals, the behavior of the thermal
conductivity can be described as $\kappa=\alpha\cdot T^n$, the
exponent $n$ and the coefficient $\alpha$ are close for both the
substances in each interval (see Table~1).

At $T=50$ K, the thermal conductivity of the orientational glass of
deuterated ethanol is about 1.3 times higher than that of protonated
ethanol. The difference can be explained qualitativelly on the
grounds of heat transfer: in this temperature interval, the hopping heat
transfer can start operating in addition to the ``diffuse'' heat
transfer. Both of the mechanisms are sensitive to the density of the
vibrational states of phonons.

\begin{figure}
\includegraphics[width=8cm]{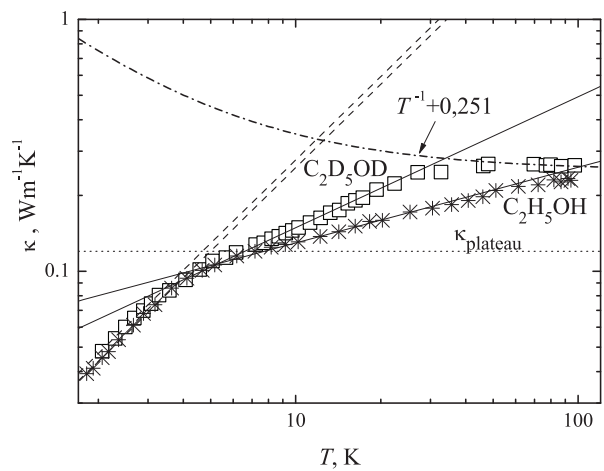}
\vskip-3mm\caption{Temperature dependence of the thermal
conductivity of orientational glasses of protonated ($\ast$) and
deuterated ($\square$) ethanol [2]. The solid and dashed straight
lines show the power law of the thermal conductivity
$\kappa=\alpha\cdot T^n$; the dotted line is the thermal
conductivity at the beginning of the ``plateau'' $\kappa_{\rm
plateau}$; the dash-dotted line is the dependence $\kappa(T)=
A/T+C$, where $A=1$~Wm$^{-1}$,
$C=0.251$~Wm$^{-1}$K$^{-1}$}\vspace*{-2mm}
\end{figure}

As the temperature increases ($T>50$~K), the thermal conductivity of
the orientational glass of the protonated alcohol (unlike the
deuterated one) is saturated. The thermal conductivity of the
orientational glass of the deuterated alcohol has a smeared maximum
in a vicinity of 50~K. At the further rise of the temperature, its
dependence can be described as $A/T+C$. This suggests that, in the high-temperature region, where the orientational glass of deuterated
alcohol is influenced by the ``diffusive'' and hopping mechanisms of
heat transfer, it also experiences the processes of phonon-phonon
scattering. Such processes are typical in orientationally ordered
crystals at temperatures above the phonon maximum of thermal
conduc\-tivity.

This behavior of the thermal conductivity in the state of
orientational glass is at variance with what is observed in
structural glasses, in which the thermal conductivity continues to
grow monotonically, as the temperature increases above the interval
of the ``plateau.''

The thermal conductivities of the orientational glasses of
cyclohexanol [4], cyanocyclohexane [3], and cyclohexene [18] are
shown in Fig. 2. The curves $\kappa(T)$ of cyclohexanol and
cyanocyclohexane behave much like $\kappa(T)$ of deuterated ethanol:
they have similar three regions. The only difference is that the
smeared maximum of the thermal conductivity is shifted to 30~K. It is
seen that, at $T<5$~K and in the smeared ``plateau'' region, where
the phonon heat transfer is operative, the thermal conductivity
behaves like that of a structural glass. The thermal
conductivities of cyanocyclohexane and cyclohexene follow the power
law $\kappa=\alpha\cdot T^n$ like that of ethanol (see Table 1).

\begin{figure} %
\vskip1mm
\includegraphics[width=8cm] {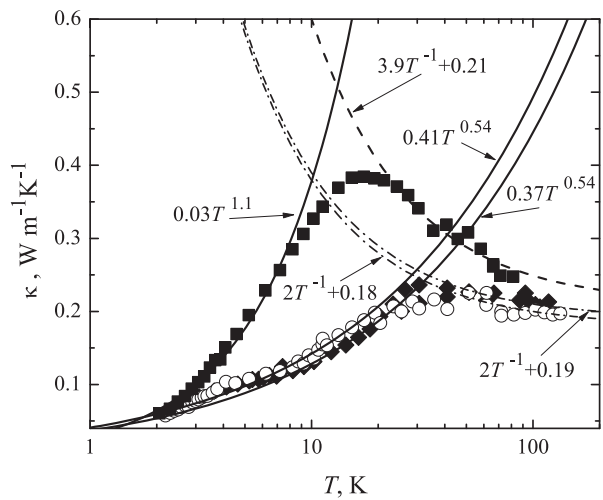}
\vskip-3mm\caption{Thermal conductivities of the orientational
glasses of cyanocyclohexane ($\circ$) [3], cyclohexanol
($\blacklozenge$) [4], and cyclohexene ($\blacksquare$) [18]. The
solid lines show the power law of the thermal conductivity $\kappa =
\alpha\cdot^T n$. The dashed and dash-dotted lines are the
dependence $\kappa(T)= A/T+C$ }
\end{figure}

\begin{table}[b]
\noindent\caption{Basic parameters of the power\\
law
\boldmath$\kappa=\alpha\cdot T^n$ describing the thermal conductivity\\
of the orientational glasses in two temperature\\ intervals $T<5$~K
and 5--50~K. $A$ and $C$\\ are the coefficients of the dependence\\
$\kappa(T)= A/T+C$, observed at $T>50$~K }\vskip3mm\tabcolsep3.0pt

\noindent{\footnotesize\begin{tabular}{|l|c|c|c|c|c|c|} \hline
\multicolumn{1}{|c}{\raisebox{-3mm}[0cm][0cm]{Samples}}&
\multicolumn{2}{|c|}{\rule{0pt}{5mm}$T<5$ K}&
\multicolumn{2}{|c|}{5--50 K}&
\multicolumn{2}{|c|}{$T>50$ K}\\[1.5mm]
\cline{2-7}& \multicolumn{1}{|c}{\rule{0pt}{5mm}$\alpha$, $\frac{W
m^{-1}}{K^{n+1}}$} & \multicolumn{1}{|c}{$n$}&
\multicolumn{1}{|c}{$\alpha$, $\frac{Wm^{-1}}{K^{n+1}}$} &
\multicolumn{1}{|c}{$n$}& \multicolumn{1}{|c}{A, $\frac{W}{m}$}&
\multicolumn{1}{|c|}{C, $\frac{W}{m \cdot K}$}\\[2mm]%
\hline%
\rule{0pt}{4mm}C$_2$H$_5$OH& 0.020& 1.1& 0.065& 0.3~\,& --& -- \\%
C$_2$D$_5$OD& 0.022& 1.1& 0.045& 0.52& 1& 0.251 \\%
C$_6$H$_{11}$OH & 0.026& 1& 0.037& 0.54& 2& 0.19~\, \\%
C$_6$H$_{11}$CN& 0.026& 1& 0.041& 0.54& 2& 0.18~\, \\%
C$_6$H$_{10}$& 0.030& 1.1& --& --& 3.9& 0.21~\, \\%
(CFCl$_2)_2$& 0.065& 0.8& 0.1& 0.12& 9.5& 0 \\[2mm]%
\hline
\end{tabular}}
\end{table}

In the smeared ``plateau'' region, the powers $n$ for cyclohexanol
and cyanocyclohexane ($n=0.54$) differ significantly from $n$ measured
for glycerol ($n=1$) [21], in which the dependence $\kappa(T)$ is
linear above the ``plateau'' temperature. The difference owes to the
rotational degrees of freedom of the orientational glass molecules
that influence the dependence $\kappa(T)$ in these cyclic
substances.

The rotational degrees of freedom may also affect the dependence
$\kappa(T)$ in the orientational glass of the cyclohexene, in which
its behavior more closely resembles that in orientationally
ordered crystals.

At low temperatures, the dependence of $\kappa(T)$ follows the law
$\kappa=\alpha\cdot T^n,$ but the power $n$ is higher than those for
the other cyclic substances. The thermal conductivity reaches its
peak at $T=17$~K and then drops, as in orientationally ordered
crystals. Note that, in the orientational glass of cyclohexene, the
dependence $\kappa(T)$ is similar to that of a completely ordered
crystal, the difference being merely quan\-titative.

Since the cyclohexene molecule is pseudospherical and has no
hydrogen bonds, its orientational disorder provides no strong
scattering, like the other investigated orientational glasses. In
the whole temperature region of the orientational glass of
cyclohexene, the temperature dependence of the thermal conductivity
resembles that of an orientationally ordered crystal with numerous
defects.

At high temperatures, the thermal conductivities of the three cyclic
substances demonstrate the great importance of the phonon-phonon
scattering. For example, above 30--40~K, the thermal conductivity of
these substances can be described by the dependence
$\kappa(T)=A/T+C$ (see Fig.~2 and Table~1).

The temperature dependence of the thermal conductivity coefficient
of freon 112 (Fig.~3) has similar three temperature regions, which
can be described by the dependences found for ethanol, cyclohexanol,
and cyanocyclohexane. The parameters are given in Table~1.\,\,Of all
the considered orientational glasses, that of freon 112 has the most
extended plateau (from 5~K to 50~K) and the lowest exponent $n$. The
plateau is between two maxima of the thermal con\-ductivity.\,\,The
low-temperature maximum is at $T=4.5$~K, and another one (a smeared
maximum) is at $T=50$~K. The behavior of the temperature dependence
of the thermal conductivity changes abruptly, when the temperature
rises above 50~K. At $T>60$~K, the thermal conductivity of freon 112
is described by the dependence typical of orienatationally ordered
crystals; $\kappa(T)=A/T+C$, $C=0$. This behavior suggests a
profound effect of the phonon-phonon scattering, which is manifested
in the coefficient $A$ accounting for the intensity of the
phonon-phonon processes (see Table 1).

The coefficient $A$ varies from 0 for protonated ethanol to 9.5 for
freon 112.\,\,The high value of $A$ in freon 112 may be attributed
to the exceptional fragility of its orientational glass, which
distinguishes it among the investigated orientational glasses [16,
17].\,\,The coefficient $A$ tends to increase with the molar mass of
orientational glass.\,\,The tendency correlates with the behavior of
the coefficient $A$ in the orientationally ordered phases of simple
alcohols, in which it increases almost linearly with the molar mass
[22].\looseness=-1

Thus, the thermal conductivity of the investigated orientational
glasses reveals a specific temperature behavior: it is glass-like
in the temperature region, where the heat is transferred by phonons,
and is crystal-like at high temperatures, where the phonon-phonon scattering
comes into play in addition to the ``diffusive'' and hopping
mechanisms of heat transfer.

The experimental results on the thermal conductivity of the
investigated orientational glasses were analyzed in terms of the
relaxation time, which was described within the soft-potential
model. In this model, the inverse phonon mean free path can be
subdivided into three components describing the processes of
resonance scattering of acoustic phonons at anharmonic defects.
The defects involve tunnel states, classical relaxation processes,
and quasilocalized low-frequency harmonic vibrations. These
processes differ only in energy. Their intensities are determined by
the dimensionless parameter $\bar{C}$, which characterizes the
binding force between the acoustic phonons and the two-level
systems and is generalized to all quasilocal excitations within
the soft-potential model. The energy $W$ is another important
parameter in this model. $W$ is the characteristic energy describing
practically all soft modes.

According to the soft-potential model, the reduced thermal
conductivity $\kappa(T)\bar{C}v/W^2$ ($v$ is the sound velocity)
as a function of the reduced temperature should be independent either
of the structure or the chemical composition of the substance, at
least at low temperatures up to the plateau region. To compare the
theoretical and experimental results, we calculated the thermal
conductivity as a function of the temperature within the
soft-potential model using the universal expression [23]
\[\kappa(T)=\frac{W^2}{\bar{C}v} F(z),\]
where $z=k_{\rm B} T/W$ is the dimensionless variable (normalized
temperature), and the function $F(z)$ is
\[F(z)=\frac{z^2 k_{\rm B}}{2\pi^2 \hbar^2} \int \limits_{0}^{\infty}
\Bigl( \frac {1}{1.1tanh(x/2)+0.7z^{3/4}+x^3 z^3/8}
\,\times\]\vspace*{-7mm}
\[\times\,\frac {x^3 e^x}{(1-e^x)^2} \Bigl)dx .\]

\begin{figure}
\vskip1mm
\includegraphics[width=8cm]{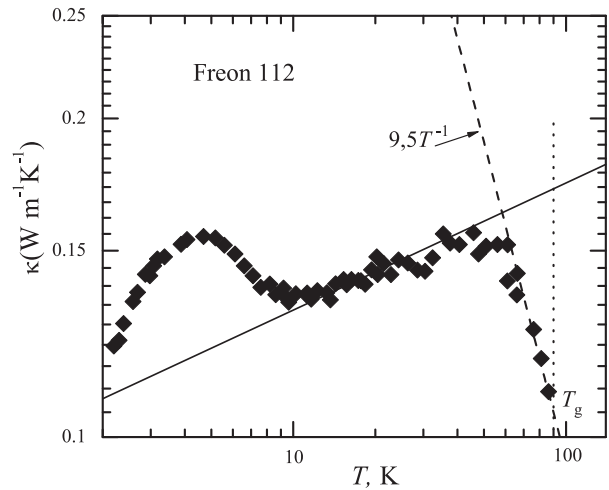}
\vskip-3mm\caption{Thermal conductivity of freon 112 in the
orientational glass state [3]. The solid straight line is the
power law $\kappa = $ $=\alpha\cdot T^n$ describing the behavior
of the thermal conductivity. The dotted line shows the glass
formation temperature $T_g=90$~K}
\end{figure}

\begin{figure}%
\vskip3mm
\includegraphics[width=8cm]{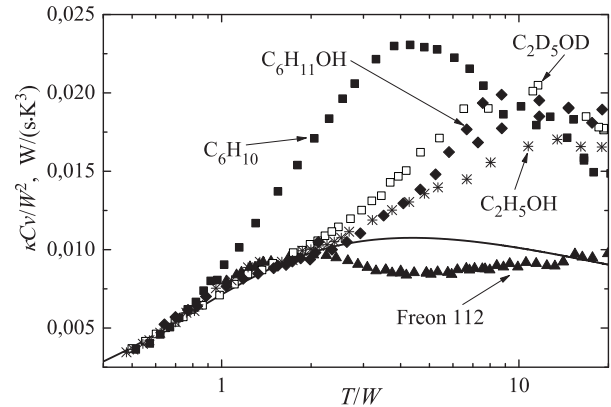}
\vskip-3mm\caption{Reduced thermal conductivity
$\kappa(T)\bar{C}v/W^2$ as a function of the reduced temperature
$T/W$. The symbols mark the experimental data on the thermal
conductivity of the orientational glasses of cyclohexanol [4],
cyclohexene [18], protonated and deuterated ethanol [2], and freon
112 [3]. The solid line is the universal dependence based on the
soft potential model}
\end{figure}

The function $F(z)$ depends only on the normalized temperature
$(z)$.

The dependence of the reduced thermal conductivity on the reduced
temperature that was obtained from the experimental results measured
on the orientational glasses of protonated and deuterated ethyl
alcohols, cyclohexene, cyclohexanol, and freon 112 is illustrated in
Fig. 4. The solid line is the unified universal curve based on the
soft potential model. It is seen that, at low temperatures $T<7.2$~K,
the experimental thermal conductivities of all orientational glasses
can be described by a single universal curve obtained within the
soft potential model.

The exception is cyclohexene, in which this temperature interval is
smaller, $T<3.3$~K. The experimental data depart from the universal
curve only in the high temperature region $T/W>2$, where the weak phonon
scattering changes to the strong one (Ioffe--Regel crossover). The
parameters of the soft potential model were estimated by fitting the
universal dependence of the function $F(z)$ to the experimental
results for the thermal conductivity of orientational glasses. The
parameters are available in Table 2.

\begin{table}[t]
\vskip4mm \noindent\caption{Parameters \boldmath$W$ and
$\bar{C}$
of the soft\\ potential model obtained by fitting the theoretical\\
curve to the experimental results taken\\ on the orientational
glasses }\vskip3mm\tabcolsep7.5pt

\noindent{\footnotesize\begin{tabular}{|l|c|c|c|c|} \hline
\multicolumn{1}{|c}{\rule{0pt}{5mm}Samples}&
\multicolumn{1}{|c}{$m$, g/mol}& \multicolumn{1}{|c}{$W$, K}&
\multicolumn{1}{|c}{$\bar{C}$, $10^{-4}$}&
\multicolumn{1}{|c|} {$T_g$, K} \\[2mm]%
\hline%
\rule{0pt}{5mm}C$_2$H$_5$OH& ~\,46.07& 3.8& 8.2& 97 \\%
C$_2$D$_5$OD& ~\,52.11& 4.1& 9.0& 97 \\%
C$_6$H$_{11}$OH& 100.16& 3.5& 4.6& 148--150 \\%
C$_6$H$_{11}$CN& 109.16& 3.5& 4.8& 135 \\%
(CFCl$_2)_2$& 203.8& 2.3& 2.8& 90 \\%
C$_6$H$_{10}$& ~\,82.14& 4~~ & 6~~ & 81 \\[2mm]%
\hline
\end{tabular}}\vspace*{-5mm}
\end{table}

In the low-temperature region, the theoretical cur\-ve based on the
soft potential model describes well the experimental results
measured on the orien\-ta\-tio\-nal glasses of the substances
consisting of the li\-near and cyclic molecules. It also describes
adequately the temperature behavior of the thermal conducti\-vi\-ty
in the orientation glasses of substances with or wi\-thout hydrogen
bonds.

In the high temperature region, when the ``diffusive'' and hopping
mechanisms are in operation, the orientational glasses also
experience the effect of the phonon-phonon scattering typical of
orientationally ordered crystals above the temperature of the phonon
maximum of the thermal conductivity.

\section{Conclusion}

The temperature dependences of the thermal conductivity of the
orientational glasses of protonated and deuterated ethanol, cyclic
cyclohexanol alcohol (including its analogs -- cyanocyclohexane and
cyclohexene), and freon 112 have been analyzed. Universal temperature
dependences of the thermal conductivity have been revealed in the
investigated orientational glasses. At low temperatures, the thermal
conductuvity exhibits a universal behavior that can be described by
the soft potential model. At high temperatures, the thermal
conductivity has a smeared maximum and then decreases with
increase in the temperature, following the law $\kappa(T)= A/T + C$, which
is common to crystalline structures and is caused by the phonon-phonon
scattering.

\vskip3mm {\it The study was jointly supported by the NAS of Ukraine
and the Russian Foundation for Basic Research Project (Agreement N
7/Н-2013). Subject: Me\-tastable states of simple condensed
systems.}

\rezume{%
О.І.\,Кривчіков,\\ Г.О.\,Вдовиченко, О.О.\,Королюк,
О.О.\,Романцова\vspace*{1mm}}{ТЕПЛОПРОВІДНІСТЬ\\ МОЛЕКУЛЯРНИХ
КРИСТАЛІВ З БЕЗЛАДДЯМ,\\ ЩО САМООРГАНІЗУЄТЬСЯ\vspace*{1mm}}
{\rule{0pt}{11pt}Проведено аналіз теплопровідності деяких
орієнтаційних стекол~-- протонованого C$_2$H$_5$OH і дейтерованого
C$_2$D$_5$OD етанолу та циклічних сполук~-- циклогексанолу
C$_6$H$_{11}$OH, цианоциклогексану C$_6$H$_{11}$CN, циклогексену
C$_6$H$_{10}$, а також фреону 112 (CFCl$_2)_2$ в температурному
інтервалі 2--130~K. Ці речовини демонструють нові ефекти, що
відносяться до фізики розупорядкованих систем. Виявлено універсальні
залежності теплопровідності молекулярних орієнтаційних стекол: при
низьких температурах спостерігається універсальна поведінка
теплопровідності, що описується моделлю м'яких потенціалів; при
відносно високих температурах спостережено розмитий максимум
теплопровідності та її зменшення з подальшим зростанням температури,
що характерно для кристалічних структур.}


\begin{thebibliography}{9}                                                                                                %

\bibitem {1}A.I.~Krivchikov, A.N.~Yushchenko, V.G.~Manzhelii {\it et al.}, Phys. Rev. B. \textbf{74}, 060201(R) (2006)

\bibitem {2}A.I.~Krivchikov, F.J.~Bermejo, I.V.~Sharapova {\it et al.}, Fiz. Nizk. Temp. \textbf{37}, 651 (2011).\vspace*{-0.3mm}

\bibitem {3}I.V.~Sharapova, A.I.~Krivchikov, O.A.~Korolyuk {\it et al.}, Phys. Rev. B. \textbf{81}, 094205 (2010).\vspace*{-0.3mm}

\bibitem {4}A.I.~Krivchikov, O.A.~Korolyuk, I.V.~Sharapova {\it et al.}, Phys. Rev. B. \textbf{85}, 014206 (2012).\vspace*{-0.3mm}

\bibitem {5}C.~Talon, M.A.~Ramos, S.~Vieira {\it et al.}, Phys. Rev. B. \textbf{58}, 745 (1998).\vspace*{-0.3mm}

\bibitem {6}M.~Mizukami, H.~Fujimori, and M.~Oguni, Solid State Comm. \textbf{100}, 83 (1996).\vspace*{-0.3mm}

\bibitem {7}D.~Ceccaldi, Phys. Rev. B. \textbf{31}, 8221 (1985).\vspace*{-0.3mm}

\bibitem {8}R.M.~Ibberson, S.~Parsons, D.R.~Allan {\it et al.}, Acta Cryst. B. \textbf{64}, 573 (2008).\vspace*{-0.3mm}

\bibitem {9}E.~Bonjour, R.~Calemczuk, R.~Lagnier {\it et al.}, J. de Phys. Colloq. \textbf{42}, C6-63 (1981).\vspace*{-0.3mm}

\bibitem {10}U.~John and K.P.R.~Nair, Spectrochim. Acta A. \textbf{61}, 2555 (2005).\vspace*{-0.3mm}

\bibitem {11}O.~Andersson, R.G.~Ross, and G.~Backstrom, Mol. Phys. \textbf{66}, 619 (1989).\vspace*{-0.3mm}

\bibitem {12}J.R.~Durig, R.M.~Ward, A.R.~Conrad {\it et al.}, Mol. Struct. \textbf{967}, 99 (2010).\vspace*{-0.3mm}

\bibitem {13}L.P.~Singh and S.S.N.~Murthy, J. Chem. Phys. \textbf{129}, 094501 (2008).\vspace*{-0.3mm}

\bibitem {14}N.V.~Surovtsev, S.V.~Adichtchev, J.~Wiedersich {\it et al.}, J.~Chem. Phys. \textbf{119}, 12399 (2003).\vspace*{-0.3mm}

\bibitem {15}K.~Kishimoto, J.J.~Pinvidic, T.~Matsuo {\it et al.}, Proc. Jpn. Acad., Ser. B: Phys. Biol. Sci. \textbf{67}, 66 (1991).\vspace*{-0.3mm}

\bibitem {16}C.A.~Angell, A.~Dworkin, P.~Figuiere {\it et al.}, J. Chem. Phys. et Phys. Chim. Biol. \textbf{82}, 773 (1985).\vspace*{-0.3mm}

\bibitem {17}L.C.~Pardo, P.~Lunkenheimer, and A.~Loidl, J. Chem. Phys. \textbf{124}, 124911 (2006).\vspace*{-0.3mm}

\bibitem {18}V.A.~Konstantinov, A.I.~Krivchikov, O.A.~Korolyuk {\it et al.}, Physica B \textbf{54}, 424 (2013).\vspace*{-0.3mm}

\bibitem {19}A.I.~Krivchikov, V.G.~Manzhelii, O.A.~Korolyuk {\it et
al.}, Phys. Chem. Chem. Phys. \textbf{7}, 728
(2005).\vspace*{-0.3mm}

\bibitem {20}A.I.~Krivchikov, B.Ya.~Gorodilov, and O.A.~Korolyuk, Instr. Exper. Techn. \textbf{48}, 417 (2005).\vspace*{-0.3mm}

\bibitem {21}C.~Talоn, Q.W. Zou, M. Ramos {\it et al.}, Phys. Rev. B \textbf{65}, 012203 (2001).\vspace*{-0.3mm}

\bibitem {22}O.A.~Korolyuk, Fiz. Nizk. Temp. \textbf{37}, 526 (2011).\vspace*{-0.3mm}

\bibitem {23}A.~Ramos and U. Buchenau, Phys. Rev. B \textbf{55}, 5749 (1997).\vspace*{3mm}

\begin{flushright}
{\footnotesize Received 22.01.09}
\end{flushright}
\end{thebibliography}
\end{document}